\documentclass[aps,twocolumn,preprintnumbers,showpacs]{revtex4}
\usepackage{graphics,graphicx,color}	% Include figure files
\usepackage{amsmath}
\newcommand{\bold}[1]{\mbox{\boldmath $#1$}}    %       bold symbol
\newcommand{\GeV}{{\rm GeV}}                    %       GeV
\newcommand{\fm}{{\rm fm}}                      %       fm
\newcommand{\eps}{{\varepsilon}}		%	epsilon

\begin{document}

%++++++++++++++++++++++ begin title and abstract +++++++++++++++++++

\preprint{LBNL-61101}

\title{Dynamical phase trajectories for relativistic nuclear collisions}

\author{I.C.~Arsene$^1$, L.V.~Bravina$^1$, W.~Cassing$^2$, 
Yu.B.~Ivanov$^3$, A.~Larionov$^{2,3}$,
J.~Randrup$^4$, V.N.~Russkikh$^3$,
V.D.~Toneev$^5$, G.~Zeeb$^6$, and D.~Zschiesche$^{6,7}$}

\affiliation{
$^1${\em Fysisk Institutt, Universitet i Oslo, Oslo, Norway
\phantom{mmmmmmmmmmmmmmmmmmmmmmi,}}
\\
$^2${\em Institut f{\"u}r Theoretische Physik, 
	Justus-Liebig-Universit{\"a}t, Gie{\ss}en, Germany
\phantom{mmmmmmmn,}}
\\
$^3${\em Kurchatov Institute, Moscow 123182, Russia
\phantom{mmmmmmmmmmmmmmmmmmmmmmmmni,}}\\
$^4${\em Nuclear Science Division, Lawrence Berkeley National Laboratory,
Berkeley, CA 94720, USA}
\\
$^5${\em Joint Institute for Nuclear Research, 141980 Dubna, Russia
\phantom{mmmmmmmmmmmmmmmmm,}}
\\
$^6${\em Institut f{\"u}r Theoretische Physik,
Johann Wolfgang Goethe-Universit{\"a}t, Frankfurt, Germany
\phantom{}}
\\
$^7${\em Instituto de F\'{i}sica, Universidade Federal do Rio de Janeiro,
%C.P. 68.528, 
21941-972 RJ, 
Brazil
\phantom{mmmmm}}
}

\date{September 15, 2006}

\begin{abstract}
Central collisions of gold nuclei are simulated by several existing models
and the central net baryon density $\rho$ and the energy density $\eps$
are extracted at successive times, for beam kinetic energies of 
5-40\,GeV per nucleon.
The resulting trajectories in the $(\rho,\eps)$ phase plane are discussed
from the perspective of experimentally exploring
the expected first-order hadronization phase transition 
with the planned FAIR at GSI or in a low-energy campaign at RHIC.
\end{abstract}


\maketitle

%+++++++++++++++++++++++ end title and abstract +++++++++++++++++++

%========================================================================
\section{Introduction}

One of the major goals of high-energy heavy-ion research 
is to explore the properties of strongly interacting matter,
particularly its phase structure \cite{goals}.
The regions of temperature and baryon density that can be accessed
depend on the collision energy.
Thus systems with a very small net baryon density but rather high temperature
are formed at RHIC \cite{Adler:2001bp} $(s_{NN}\simeq(200~\GeV)^2)$,
while it is expected that the creation of the 
highest possible baryon densities 
would occur at more moderate collision energies 
($s_{NN}\simeq(6~\GeV)^2$ \cite{IvanovPRC73}),
such as those becoming available at the planned FAIR \cite{FAIR}
or at the low-energy end of RHIC.

Our understanding of the QCD phase diagram is best developed
at vanishing chemical potential, $\mu_B=0$,
where lattice QCD calculations are most easily carried out.
The most recent results indicate that the transformation
from a low-entropy hadron resonance gas
to a high-entropy quark-gluon plasma
occurs smoothly as the temperature is raised,
with no real phase transition being present \cite{karsch_qm}.

On the other hand, at zero temperature
most models predict the occurrence of a first-order phase transition 
when the density is raised \cite{Stephanov:1998dy},
though no firm results are yet available
for the corresponding value of the chemical potential, $\mu_0$.
However, if the $T=0$ transformation is in fact of first order,
one would expect the phase boundary 
to extend into the region of finite temperature and terminate 
at a certain critical endpoint, $(\mu_c,T_c)$ \cite{Stephanov:1998dy}.
Indeed, recent lattice QCD results \cite{fodor} suggest the presence 
of such a first-order phase transition line
and an associated critical end-point,
though its precise location is not yet determined.

The Facility for Antiproton and Ion Research (FAIR) 
under construction at GSI in Germany
will make it possible to create compressed baryonic matter in the laboratory,
matter with a high net baryon density.
The increasing interest in this physics area is underscored by the recent
proposal for a low-energy campaign at RHIC
aimed at the identification of the critical point \cite{le-RHIC}
and by current discussions about the feasibility of searching for 
the mixed phase at the JINR Nuclotron \cite{Nuclotron}.

In order to assess the prospects for using these facilities to explore the
phase structure it is important to know
what thermodynamic environments are being generated in the bulk of the
collision systems at the various bombarding energies available.
For this purpose, we have employed a number of existing models
to simulate central collisions of gold nuclei 
in the beam energy range anticipated at FAIR ($5-40~\GeV/A$)
and then extracted key information about
the bulk environments generated in the course of a collision.

%========================================================================
\section{The information extracted}

Generally, the systems involved in a high-energy nuclear collision
evolve rapidly in time and, furthermore, 
they are far from being uniform in space.
The former feature prevents equilibrium from being fully established,
while the latter feature invalidates the familiar thermodynamic relations
which pertain to bulk matter.
As a consequence, dynamical simulation models are indispensable
in the exploration of these processes.

However, within a given microscopic transport model,
it is possible to extract the characteristics of the local environment
at any point in space and time and on this basis ascertain the degree of
local equilibrium achieved and extract the corresponding local characteristics.

Such a study was first made by Dorso {\em et al.}\ \cite{DorsoPLB232}
who calculated the breakup of an initially compressed and heated nucleus
and extracted its thermodynamic phase evolution.
This analysis showed that the bulk of the system entered the spinodal region 
associated with the first-order nuclear liquid-gas phase transition.
Furthermore, 
the resulting fragmentation pattern exhibited signs of filamentation,
a general characteristic feature of spinodal phase decomposition. 

As an instructive reference case for our present study,
we consider central collisions of two gold nuclei
and focus on the physical conditions at the center of the system.
Thus, in the CM frame, we consider only a small region around the origin,
$\bold{r}=(0,0,0)$, and then seek to characterize the physical environment 
there as it evolves in the course of time.

We are particularly interested in the net baryon density 
$\rho(t)=\rho_B(t)-\rho_{\bar B}(t)$ and the energy density $\eps(t)$.
The local stress tensor is also of interest
but will not be examined here.
Since we focus on the center of the system,
there is no collective flow by symmetry.
(While this is strictly true only on the average,
each individual event might display some flow at the origin,
but this possibility is unimportant and may be safely disregarded.)
We do not wish to engage in a technical discussion of how these
quantities can be extracted in the various models
but refer the reader to the relevant literature cited 
for each particular model.
We only note here that 
in numerical treatments that employ a grid in position space,
such as fluid dynamics, these values can simply be read off
at the appropriate lattice site (the origin, in the present case),
while methods that represent the dynamical state of the system
in terms of individual (test) particles
must resort to an average over a suitably small test volume $\Delta V$
around the origin.
[Due to the strong Lorentz contraction early on,
the test volume must initially be sufficiently thin;
a typical choice would be $|x|, |y|, \gamma_{\rm cm}|z|\leq2\,\fm$,
where $\gamma_{\rm cm}$ is the Lorentz factor associated with 
the initial nuclear motion in the CM frame.
On the other hand, as the longitudinal expansion progresses
and the system grows increasingly dilute,
it may improve the sampling statistics to stretch the test box.]

It is important to note that both $\rho$ and $\eps$ have well-defined values
at all times.
In particular, their extraction does not require that
local thermal equilibration has been reached.
This is one advantage of considering these particular observables
for the present study.
However, of course, their thermodynamic relevance does depend on 
the degree of local equilibration achieved,
as reflected principally in the isotropy of the pressure tensor.

In each individual model, 
it may be possible to also extract local thermodynamic quantities, 
such as temperature $T$, chemical potential $\mu$, 
or entropy density $\sigma$,
but although most extraction methods can be cast in sufficiently general
terms to make them applicable also to non-equilibrium scenarios,
those quantities have physical meaning only in equilibrium.
Furthermore, importantly, even if equilibrium is reached,
identical values of $\rho$ and $\eps$ will generally lead to different
values for those thermodynamic quantities from one model to the other,
due to their differences in mean fields and degrees of freedom.

By contrast, the mechanical quantities $\rho$ and $\eps$ 
are inherently more robust variables
since they are subject to local conservation laws.
For example, in ideal fluid dynamics 
the conservation of four-momentum is expressed as $\partial_\mu T^{\mu\nu}=0$,
while the conservation of baryon charge is expressed by the continuity
equation $\partial_\mu j^\mu=0$.
Since the various dynamical models generally abide by these basic
conservation laws, they will have a tendency to yield similar results
for the corresponding quantities.
By asking about the behavior of such conserved observables
we may therefore expect to obtain relatively robust answers.
[Of course, for the purpose of discriminating between models
(which is {\em not} our purpose here),
it would probably be better to consider observables that are more sensitive
to the specific model ingredients.]

To further underscore the qualitative difference between
``mechanical'' or ``dynamical'' quantities such as $\rho$ and $\eps$
and ``thermodynamical'' quantities such as $\mu$ and $T$,
we note that the above-mentioned conservation laws guarantee
that the local energy or (baryon) charge density cannot change
without the occurrence of a suitable amount of energy or charge transport,
which requires some time.
By contrast, there are no such conservation laws restricting the
rate of change of the local temperature or entropy or chemical potential,
which can change essentially instantaneously
as a result of local reaction processes,
such as ionization or chemical bonding.
[While such phenomena may well offer useful signals of the
hadronization phase transition (which can be viewed as some sort of bonding),
they are not of interest for the present study.]

A further advantage of considering the variables $\eps$ and $\rho$
rather than $T$ and $\mu$ is that the equation of state,
{\em i.e.}\ the pressure $p(\eps,\rho)$ is then always a single-valued function
while this is not always the case for $p(T,\mu)$.
Indeed, precisely when a first-order phase transition is present,
the bulk pressure ({\em i.e.}\ the pressure of a spatially uniform system),
$p(T,\mu)$, is multi-valued throughout the region of phase coexistence.
Because of this feature, if the phase trajectory were represented as
$(\mu(t),T(t))$, it would exhibit a rather complex behavior 
as the expansion drives the system through the phase coexistence region, 
thus complicating the analysis considerably.
This problem is not encountered in the $(\rho(t),\eps(t))$ representation,
where the phase trajectory has a regular behavior throughout.
This makes it easy, for example, to see how long time will be spent in the 
spinodal phase-coexistence region where bulk matter is mechanically unstable.

%========================================================================
\section{Phase diagram}
\label{diagram}

\begin{figure}          %       -----------------------------------------
\includegraphics[angle=0,width=3.1in]{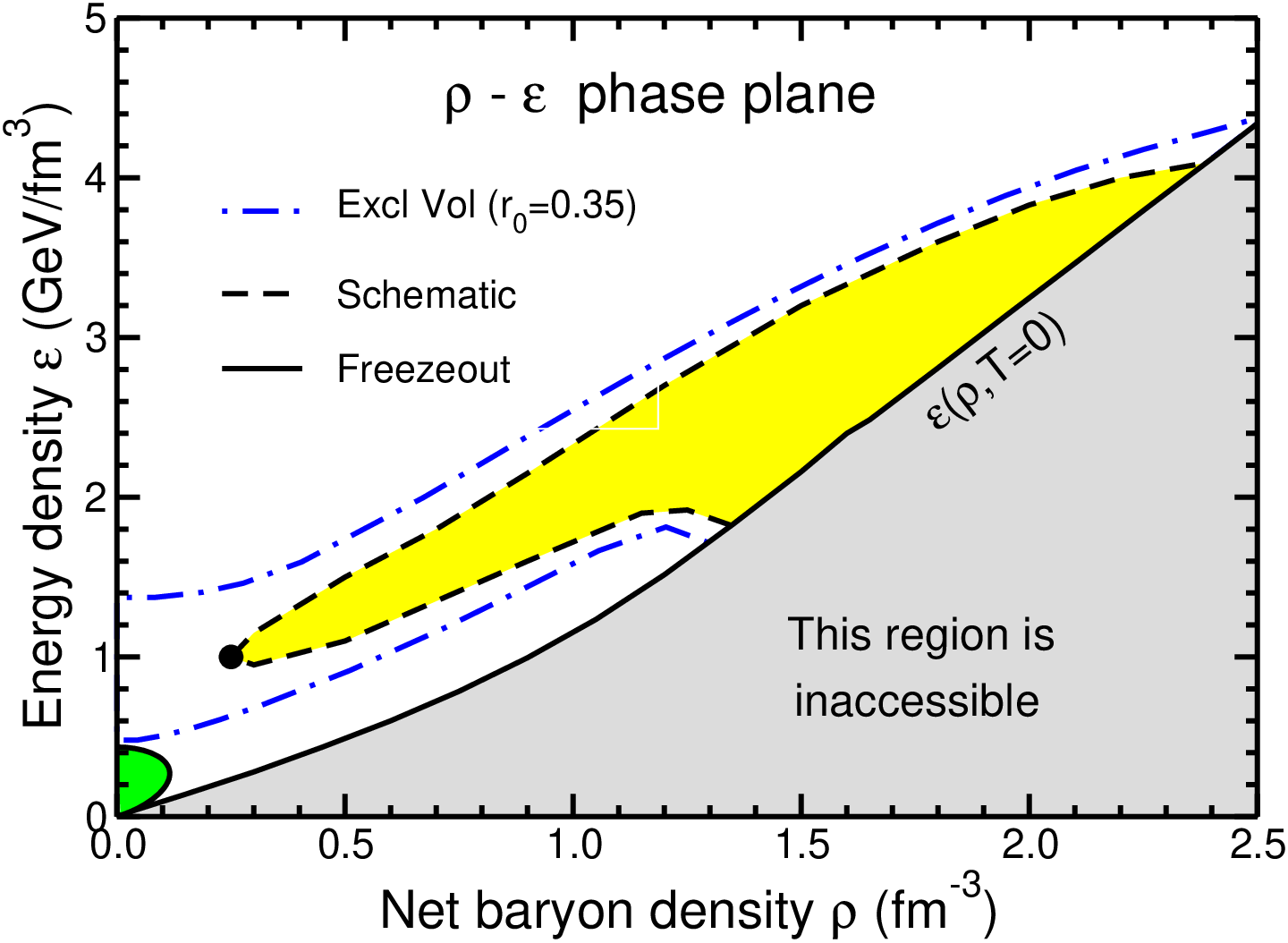}
\caption{[Color online]
The $\rho$-$\eps$ phase plane used for representing the extracted
dynamical evolution of the central environment in Au+Au collisions.
The energetically accessible region is bounded from below 
by the zero-temperature compressional energy density $\eps_{T=0}(\rho)$.
The hadronic freezeout is indicated at the lower left. % (green).
The phase coexistence region obtained in Ref.\ \cite{EoS}
on the basis of an excluded volume is delineated by the outer contour,
while the inner contour, % (yellow), 
which depicts a schematic boundary with a critial point, 
will serve as a reference for the phase trajectories.
}\label{f:EoS}
\end{figure}            %       -----------------------------------------

The most relevant features of the $\rho-\eps$ phase plane 
are depicted in Fig.\ \ref{f:EoS}.
At a given density $\rho$, the zero-temperature compressional energy,
$\eps_{T=0}(\rho)$,
provides a lower bound on the energy density $\eps$,
so the accessible region is correspondingly limited.
A useful reference is provided by the phase-coexistence boundaries 
associated with a recently constructed equation of state \cite{EoS}
that has a first-order phase transition at all baryon densities.
Also shown is the corresponding boundary of a schematic equation of state
that has a critical point at a finite density, as is now generally expected.
While these boundaries are only approximate and illustrative,
they may serve as convenient references
on the plots of the calculated phase trajectories.

Figure \ref{f:EoS} also shows where the hadronic freeze-out occurs
\cite{freezeout}.
This representation brings out the fact that the freeze-out environments
are quite different from those near the phase coexistence boundary,
thus underscoring the importance of studying the propagation
and survival of any proposed phase-transition signals through freeze-out.

%-----------------------------------------------------------------------
\subsection{Isentropic expansion}
\label{SU3}

\begin{figure}[b]          %       -----------------------------------------
\includegraphics[angle=0,width=3.1in]{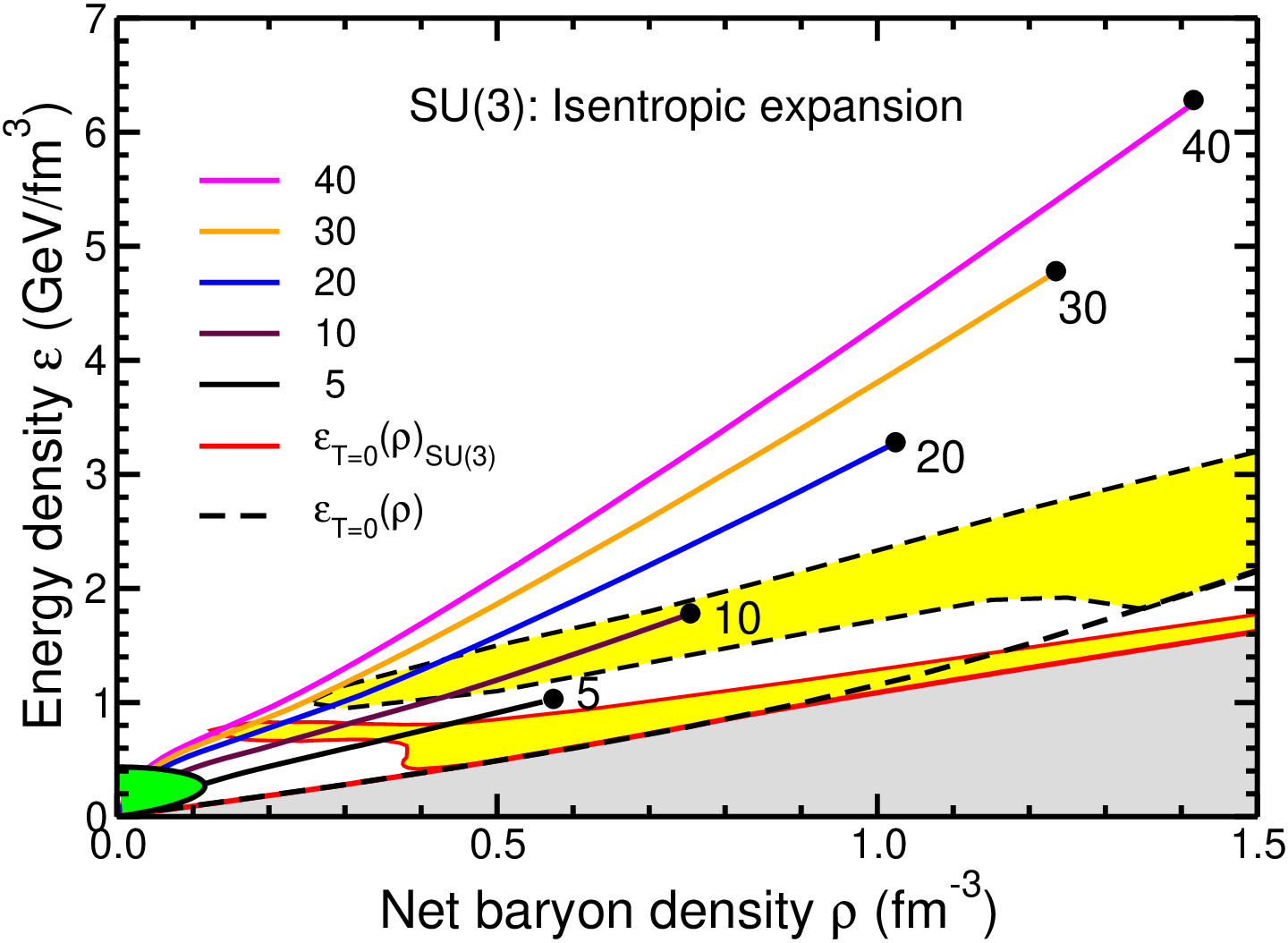}
\caption{[Color online] The chiral hadronic SU(3) model:
The energetically inaccessible region (grey area)
and the phase-coexistence region (shaded area within solid boundary)
are shown together with the corresponding quantities for the
schematic equation of state displayed in Fig.\ \ref{f:EoS}.
Also shown are the phase trajectories $(\rho,\eps)$ resulting from
adiabatic expansions starting from counterstreaming Lorentz-contracted nuclei
at the indicated beam kinetic energies $E_0$ (GeV/$A$).
}\label{f:SU3}
\end{figure}            %       -----------------------------------------

In order to establish an instructive framework for understanding
the results obtained with the various dynamical models, 
we consider first adiabatic expansions.
For this we use the hadronic chiral flavor-SU(3) model \cite{PapazoglouPRC59}.
This model is based on a chiral hadronic SU(3) Lagrangian that incorporates 
and couples the complete set of baryons from the lowest flavor-SU(3) octet,
the entire multiplets of scalar, pseudoscalar, vector, 
and axial vector mesons, as well as baryon resonance states
 \cite{PapazoglouPRC59,ZschieschePRC65,ZschiescheJPG31}.
These hadrons have various types of interaction
that endow them with effective masses
and induce spontaneous chiral symmetry breaking
as well as scale breaking via a dilaton field.
The parameters of the model are fixed by symmetry considerations,
hadronic vacuum observables, or nuclear matter saturation properties.
The model provides a satisfactory description of both finite nuclei
and neutron stars \cite{PapazoglouPRC59,SchrammPRC66,SchrammPLB560}
and, furthermore, it has been used for fluid-dynamical studies 
of the space-time evolution and HBT radii
in relativistic nuclear collisions \cite{ZschieschePRC65}.

With the baryon resonance couplings chosen suitably,
the phase diagram of the SU(3) model is in qualitative agreement 
with the picture obtained from lattice results \cite{SU3},
as illustrated by the corresponding phase boundary in Fig.\ \ref{f:SU3}.
(But it is seen to differ quantitatively from the schematic reference boundary
discussed above - a useful reminder of that fact that the phase boundary
is still rather poorly understood.)
Furthermore,  phase trajectories that are consistent with the phase diagram 
of the model can be obtained by performing adiabatic expansions.
Such expansions conserve the entropy per net baryon and that condition
in turn yields a unique trajectory in the $\rho-\eps$ phase plane,
once the initial phase point $(\rho_i,\eps_i)$ has been specified.

A simple but rough estimate of the initial conditions 
can be obtained by assuming that the very early dynamics 
is dominated by the interpenetration of the two Lorentz-contracted nuclei.
Then the early baryon density (in the CM frame) is 
$\rho_i=2\gamma_{\rm cm}\rho_0$,
where $\rho_0\approx0.15~\fm^{-3}$ is the normal nuclear saturation density
present in the nuclear interior and $\gamma_{\rm cm}$ is the Lorentz factor
of the nuclei in the CM frame, $\gamma_{\rm cm}^2=1+E_0/2m_N$,
where $E_0$ is the beam kinetic energy per nucleon for a stationary target.
In the CM, the energy per baryon is $\gamma_{\rm cm}m_N$,
so the energy density is 
$\eps_i=\gamma_{\rm cm}m_N\rho_i=2\gamma_{\rm cm}^2m_N\rho_0=(2m_N+E_0)\rho_0$.
The resulting phase trajectories $(\rho,\eps)$ 
are depicted in Fig.\ \ref{f:SU3}.
They are straight lines through the phase-coexistence region,
while they are slightly convex above it and slightly concave below it.

As we shall see, the corresponding adiabatic compression
(obtained by following these phase trajectories in the opposite direction)
are remarkably similar to the calculated dynamical paths
through the early non-equilibrium stage when counterstreaming dominates,
while the subsequent dynamical expansion trajectories 
generally exhibit gentler slopes.

%========================================================================
\section{Dynamical results}

We have employed a number of different dynamical models 
in this comparitive study.
Since they have been described in the literature already
we present only brief characterizations here
and concentrate on the resulting phase trajectories.

%-----------------------------------------------------------------------
\subsection{Three-fluid hydrodynamics}

\begin{figure}          %       -----------------------------------------
\includegraphics[angle=0,width=3.1in]{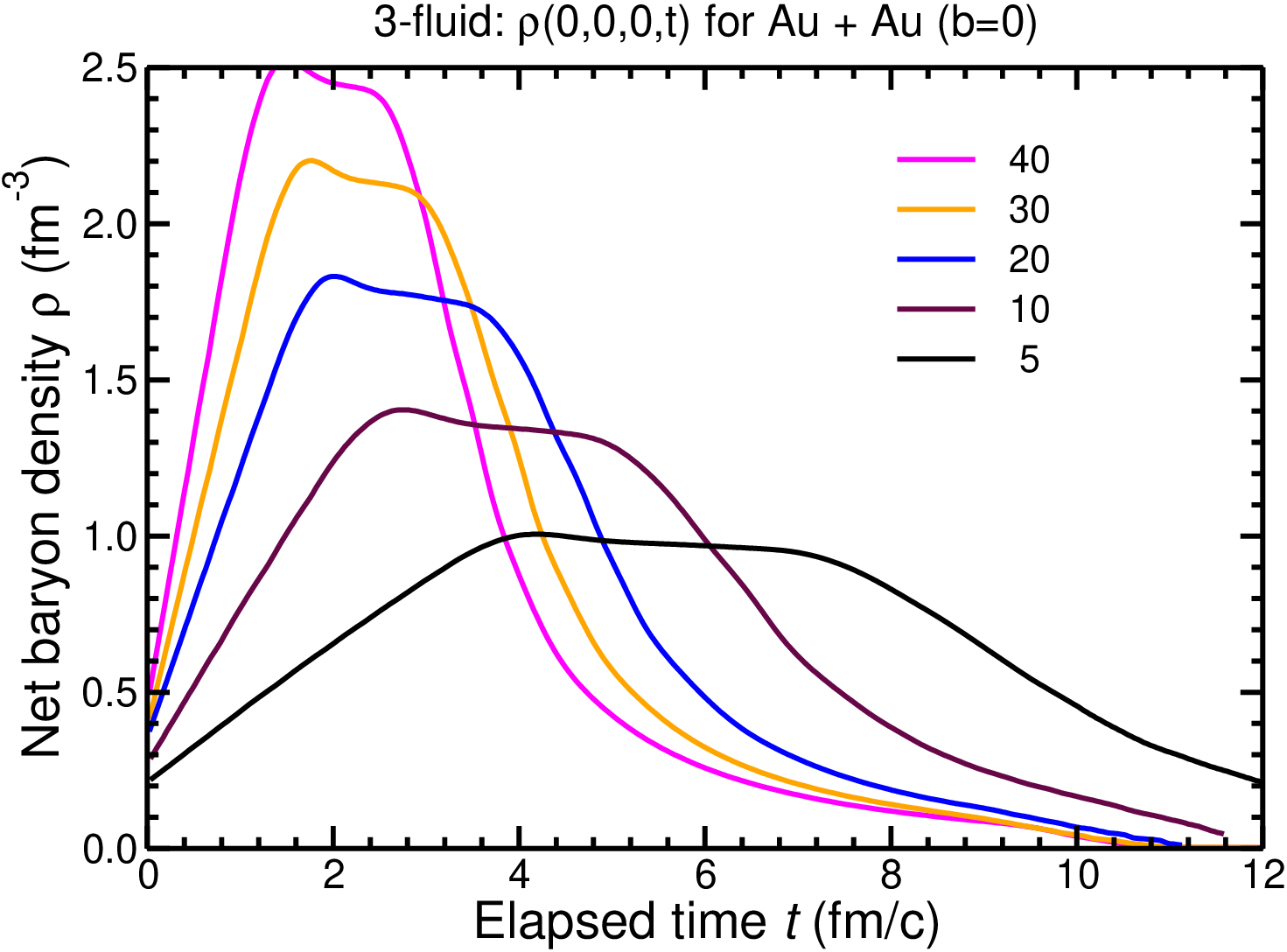}
\includegraphics[angle=0,width=3.1in]{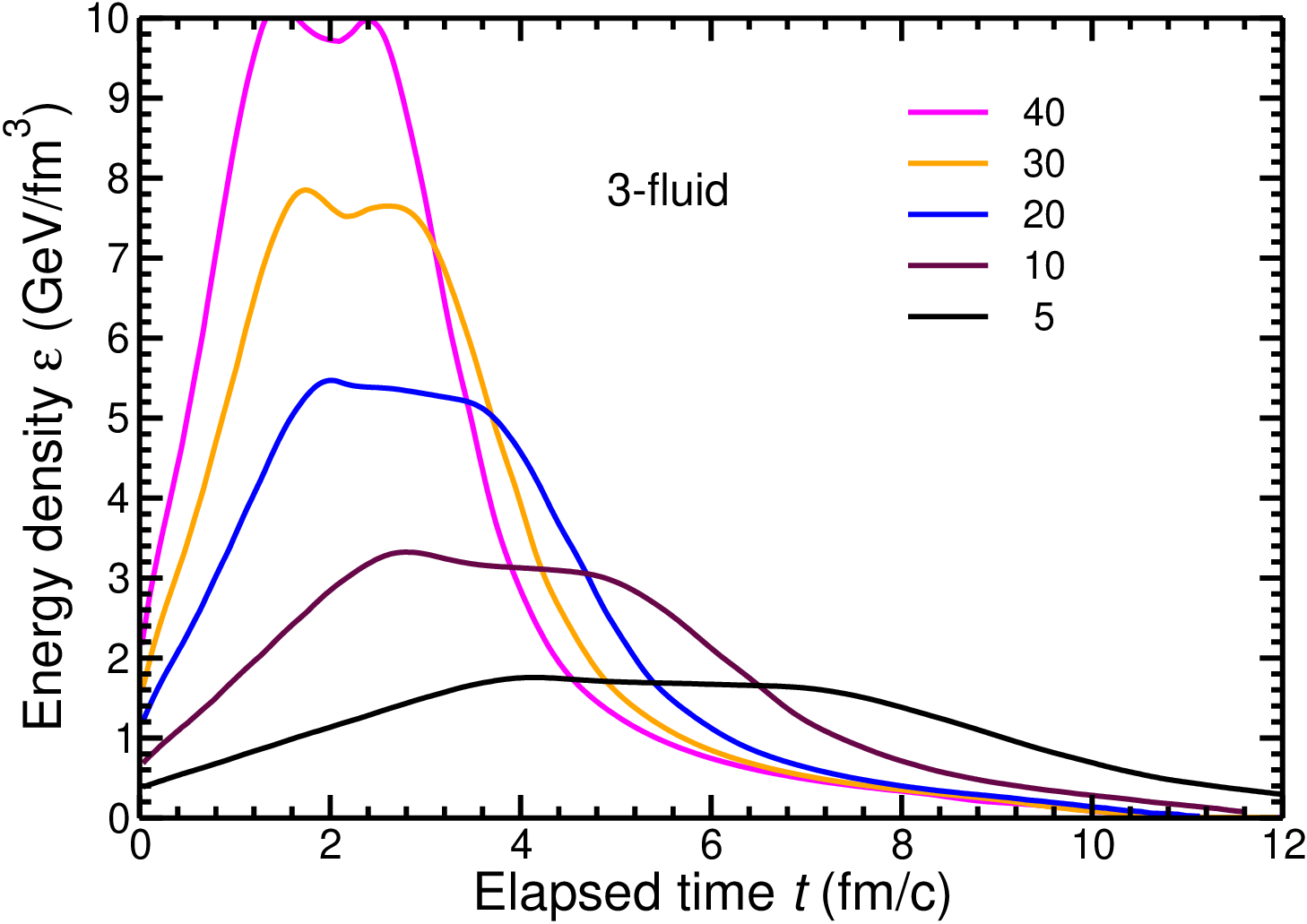}
\caption{[Color online] 
The time evolution of the net baryon density $\rho(t)$ (top)
and the total energy density $\eps(t)$ (bottom)
at the center of a head-on Au+Au collision 
for various bombarding energies (indicated in GeV/$A$), in the 3-fluid model.
}\label{f:3Fa}
\end{figure}            %       -----------------------------------------

We first consider the three-fluid model \cite{IvanovPRC73} 
which treats two baryon-rich fluids originating with the incoming nuclei 
and a baryon-free fluid created through the collisions 
among the (EoS-dependent) constituents of the first two fluids.
The evolution of the baryon-free fluid is delayed by a formation time $\tau$, 
during which it neither thermalizes nor interacts with the baryon-rich fluids.
After its formation, it starts to interact with the baryon-rich fluids 
and quickly thermalizes. 
With a purely hadronic equation of state, this model was used to carry out
a systematic analysis of various observables 
at incident energies between few and about $160~\GeV/A$,
and a comparison with results of transport models has been made as well. 

A large body of data has been well reproduced,
including proton and pion rapidity distributions, 
proton transverse-mass spectra, 
$\Lambda$ and $\bar{\Lambda}$ rapidity distributions, 
protons and pion elliptic flow
(except for the proton $v_2$ at $40~\GeV/A$), 
multiplicities of pions, positive kaons, $\phi$ mesons, hyperons, 
and antihyperons, including multistrange particles. 
This agreement is achieved at the expense of substantial enhancement 
of the interflow friction, 
as compared to that estimated from free hadronic cross sections.
Problems were met in reproducing the transverse flow \cite{Ivanov:2state},
{\em e.g.}\ the directed flow requires a softer EoS 
at top AGS and SPS energies. 
This failure appears to suggest that the employed 
purely hadronic equation of state is too hard 
and thus leaves room for softening due to deconfinement.
Further studies are in progress \cite{Ivanov:mt}.

The evolutions of the central values of $\rho$ and $\eps$ 
in head-on Au+Au collisions obtained with the three-fluid model
are depicted in Fig.\ \ref{f:3Fa} 
for beam kinetic energies ranging from 5 to 40 $\GeV/A$.
At each beam energy, $\rho(t)$ and $\eps(t)$ are approximately proportional
and exhibit a rapid growth as the two Lorentz-contracted nuclei interpenetrate,
followed by a somewhat slower decrease reflecting the subsequent expansion.
As the beam energy is increased, 
the entire history is being compressed in time.

\begin{figure}          %       -----------------------------------------
\includegraphics[angle=0,width=3.1in]{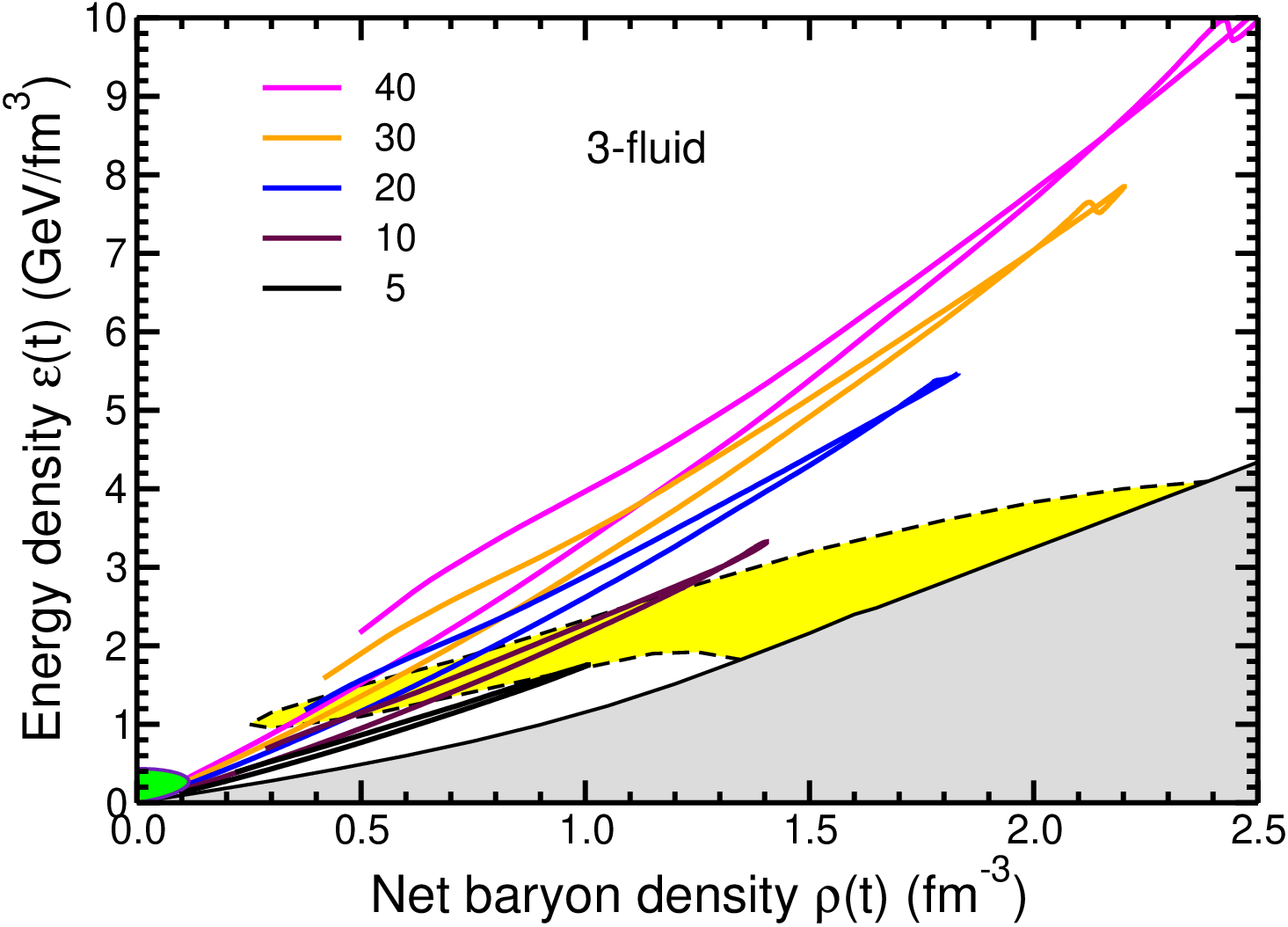}
\caption{[Color online] The phase trajectories $(\rho(t),\eps(t))$
for the 3-fluid collisions addressed in Fig.\ \ref{f:3Fa},
together with the schematic reference phase boundary 
depicted in Fig.\ \ref{f:EoS}.
}\label{f:3F}
\end{figure}            %       -----------------------------------------

The separate evolutions $\rho(t)$ and $\eps(t)$ are then combined
in Fig.\ \ref{f:3F} to yield the corresponding dynamical phase trajectory
$(\rho(t),\eps(t))$.
At each beam energy, the return path depicting the expansion
lies below the early (outwards) path,
although the two paths generally differ only relatively little.
As was the case for the adiabatic results considered above,
each such ``common'' path is fairly straight and
its slope increases steadily with the beam energy.
For the lowest energy, $5~\GeV/A$, the trajectory just makes it
to the hadronic boundary of the schematic phase coexistence region,
while the next energy, $10~\GeV/A$, already produces a turning point
on the plasma side beyond the schematic phase coexistence region.
Thus, of the trajectories shown, this one spends the longest time
traversing the phase coexistence region during the expansion phase,
and the crossing time becomes ever shorter as the beam energy is raised.

It should be recalled that the projection onto the $(\rho,\eps)$
phase plane does not require any assumption about local equilibrium
(which is very convenient for the purpose of the present study).
In fact, generally, the central conditions are far from equilibrated
during the early part of the collision.
Therefore, before any thermodynamical implications could be made
it would be necessary to carefully analyze the degree of equilibrium attained
at any particular time of interest.

%-----------------------------------------------------------------------
\subsection{Parton-hadron string dynamics}

We now consider a number of microscopic transport models. 
The first one is a recently extended version of the HSD model 
\cite{CassingNPA674} 
called PHSD (Parton-Hadron-String-Dynamics) \cite{PHSD}. 
This version includes additionally an early partonic phase 
with an equation of state from lattice QCD 
and quasi-particle properties for quarks, antiquarks and gluons 
that have been obtained from fits to lattice results \cite{PershierPRL94}. 
On the hadronic side it treats explicitly the familiar baryon octet and
decouplet and selected higher resonances as well as their antiparticles.
On the meson side it includes the pseudo-scalar and vector meson nonets 
as well as some higher meson resonances ($a_1${\em  etc.}). 
Hadrons of even higher mass are treated as ``strings'' 
that reflect the continuum excitation spectrum of mesons or baryons.
Since the results from the novel PHSD are very similar to those from HSD 
(without partonic phase) for central Au+Au collisions 
below about $25~\GeV/A$ 
we omit a more detailed description of the PHSD model here. 

We recall that the HSD model has been extensively compared 
to experimental data from nucleus-nucleus reactions for energies
of $1~\GeV/A$ to the top RHIC energies ($s_{NN} =( 200~\GeV)^2$)
for hadrons made up from the light $u,d$ quarks, strange hadrons
\cite{brat03,brat04} as well as open and hidden charm \cite{brat05}.
It compares rather well with data (and the UrQMD model described below)
for observables such as hadron rapidity distributions \cite{brat03} 
but falls somewhat low in the $K^+/\pi^+$ ratio at top AGS or FAIR energies. 
Collective flow observables ($v_1(y),v_2(y)$) are reproduced rather well 
in the SIS/AGS energy regime \cite{Sahu} 
due to momentum-dependent scalar and vector selfenergies for the baryons. 
However, the transverse slope of kaons and antikaons is underestimated 
for bombarding energies above about $5~\GeV/A$ in central Au+Au collisions 
which has lead to the suggestion that a 'new phase of matter' 
or 'partonic degrees of freedom' should already be encountered 
at top AGS energies \cite{PRLbrat}.

\begin{figure}          %       -----------------------------------------
\includegraphics[angle=0,width=3.1in]{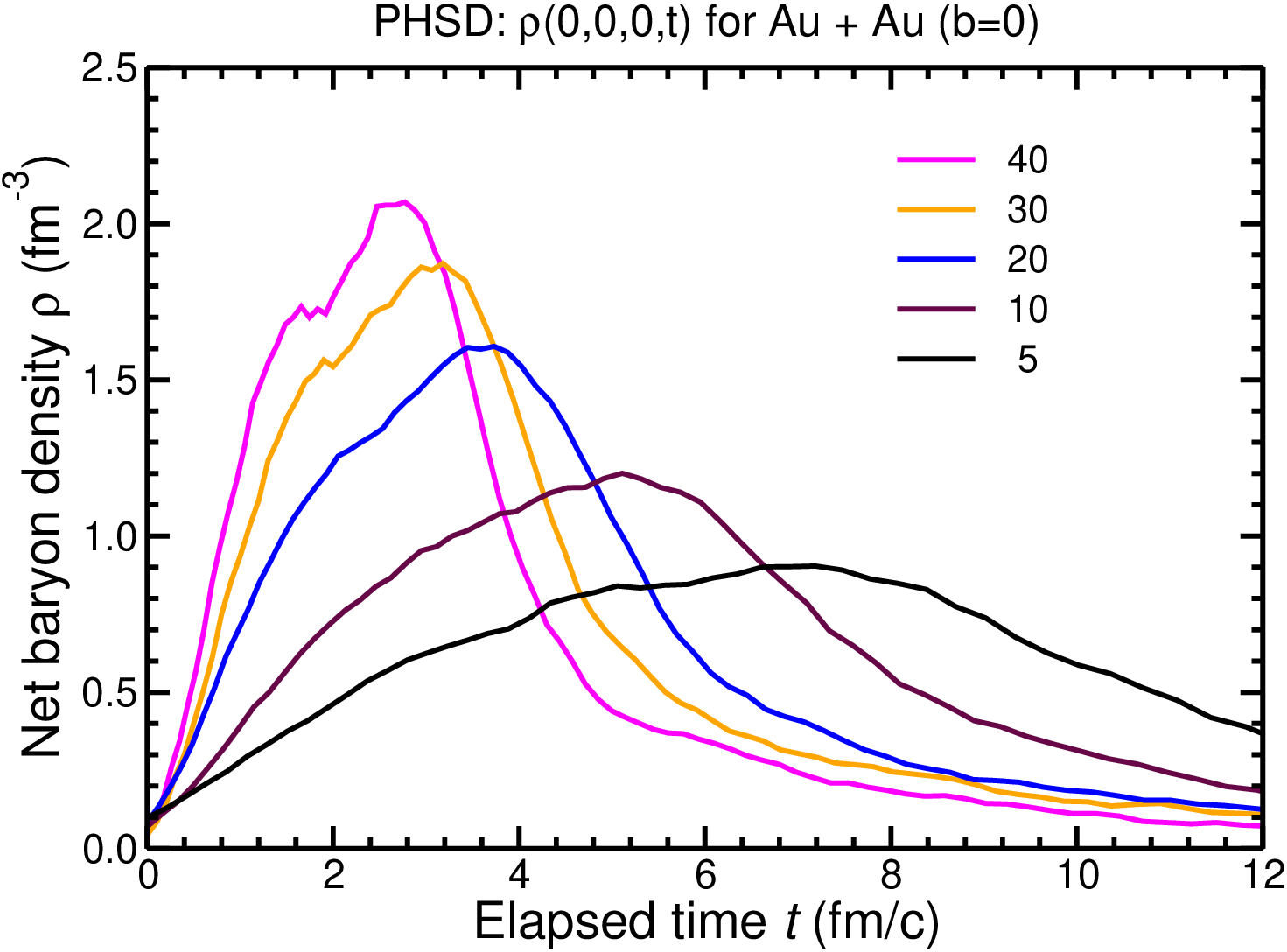}
\includegraphics[angle=0,width=3.1in]{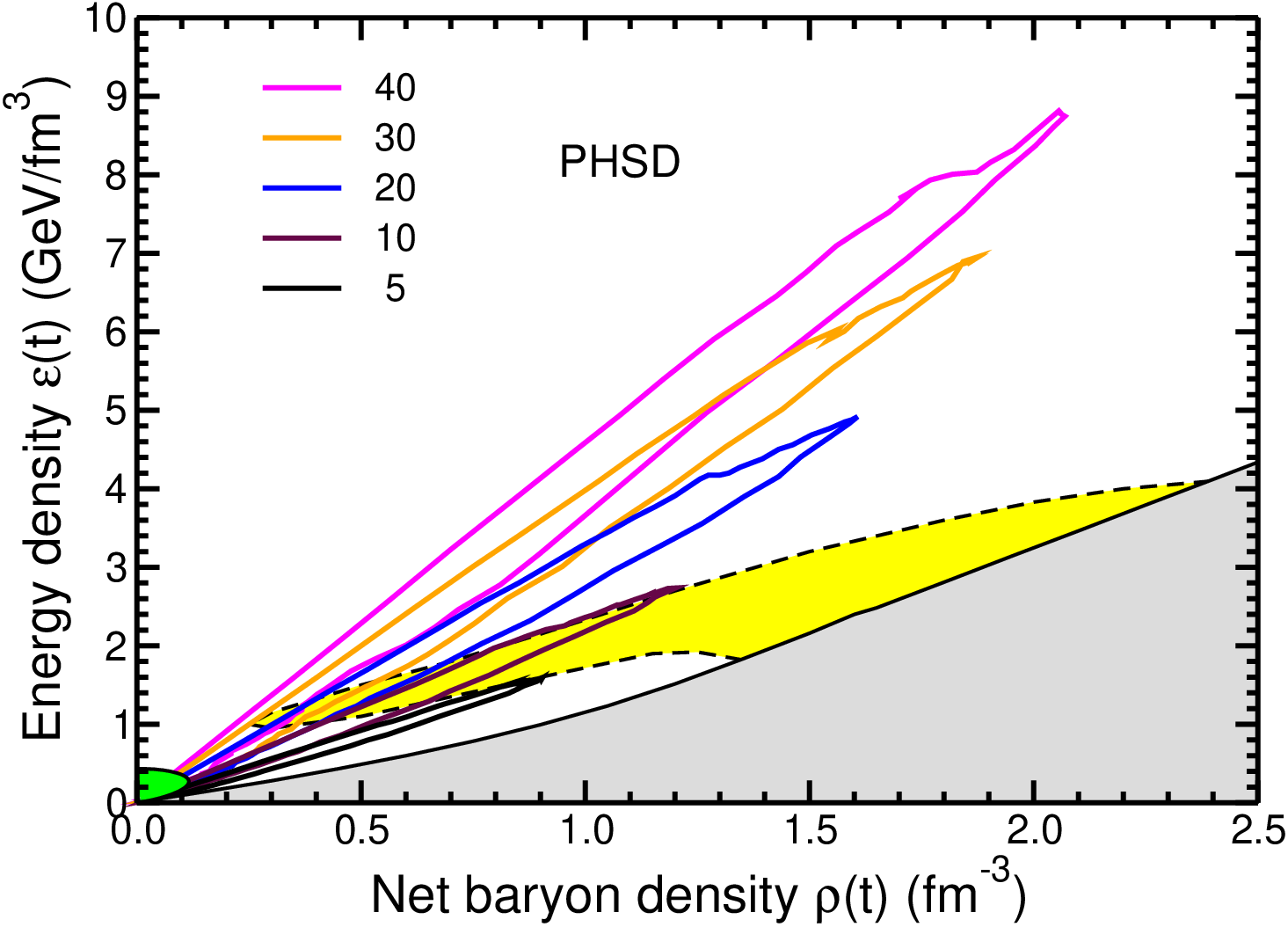}
\caption{[Color online] 
The time evolution of the central net baryon density $\rho(t)$ (top)
and the corresponding phase trajectory $(\rho(t),\eps(t))$ (bottom)
at the center of a head-on Au+Au collision at various bombarding energies
(indicated in GeV/$A$), in the PHSD model,
together with the schematic reference phase boundary 
depicted in Fig.\ \ref{f:EoS}.
}\label{f:PHSD}
\end{figure}            %       -----------------------------------------

The results obtained with PHSD are shown in Fig.\ \ref{f:PHSD}.
(Here and for the subsequent models,
we do not show $\eps(t)$ since it is approximately proportional to $\rho(t)$.)
The time evolution of the densities $\rho(t)$ are remarkably similar 
to those obtained with the 3-fluid model discussed above,
but it can be seen that PHSD leads to somewhat smaller compressions,
particularly at higher collision energies.

The PHSD phase trajectories are therefore also rather similar 
to those obtained with the 3-fluid model,
except for somewhat smaller compressions and excitations
at the highest energies.
(We may also note that both models yield a curious double-hump structure
of the density maxima, particularly at the higher energies.)
Furthermore, at each energy,
the inwards path is rather similar to the outwards path,
though the differences are larger than those obtained 
with the three-fluid model.
We also note that the $5~\GeV/A$ phase trajectory turns around
just at the hadronic border of the schematic phase coexistence region
(as for the three-fluid model),
while the $10~\GeV/A$ phase trajectory turns around
at just the schematic plasma phase boundary of this region.

%-----------------------------------------------------------------------
\subsection{Ultra-relativistic quantum molecular dynamics}

The Ultra-relativistic Quantum Molecular Dynamics model \cite{urqmd1,urqmd2}
is a microscopic model used to simulate (ultra)relativistic heavy-ion 
collisions in the energy range from BEVALAC and SIS up to AGS, SPS and RHIC
allowing for a consistent calculation of excitation functions.
Its main goals are to gain understanding of the various physical phenomena 
within a single transport model, 
including creation of dense hadronic matter at high temperatures, 
properties of nuclear matter, $\Delta$ and resonance matter, 
mesonic matter and of anti-matter,
creation and transport of rare particles in hadronic matter, creation,
modification and destruction of strangeness in matter, 
and emission of electromagnetic probes.

The initial projectile and target nuclei are modeled according 
to the Fermi gas ansatz and the nucleons are represented
by gaussian shaped density distributions. 
UrQMD includes in its collision term 55 different baryon species with masses
up to 2.25~GeV, 32 meson species (including strange meson resonances) 
which are supplemented by their corresponding anti-particle 
and all isospin-projected states.
All these states can be produced in string decays, 
$s$-channel collisions or resonance decays. 
For excitations with masses higher that 2~GeV a string picture is used.
The hadron-hadron collisions are performed stochastically like
in the cascade models. The elementary cross-sections are fitted
to available $pp$, $\pi p$ data and the isospin symmetry
is used whenever possible in order to reduce the number of
individual cross-sections. 
For the interactions where no experimental data exist 
({\em e.g.}\ hyperon-baryon resonance scattering),
the additive quark model is used.

The interactions are based on a 
non-relativistic density-dependent Skyrme-type equation of state 
with additional Yukawa and Coulomb potentials at low energies.
However, no potentials were used in the present calculations.
For the high energy regime and baryon-antibaryon annihilation,
UrQMD uses a string model similar to the Lund model \cite{Lund1,Lund2}.
The strings, or the color tubes are first formed in the high energy 
$hh$ interactions and then fragment into hadrons and new strings according 
to the Lund fragmentation procedure. 
For the ultra-high energies (top SPS energies and beyond) 
the formation of jets is also introduced into the model.

\begin{figure}          %       -----------------------------------------
\includegraphics[angle=0,width=3.1in]{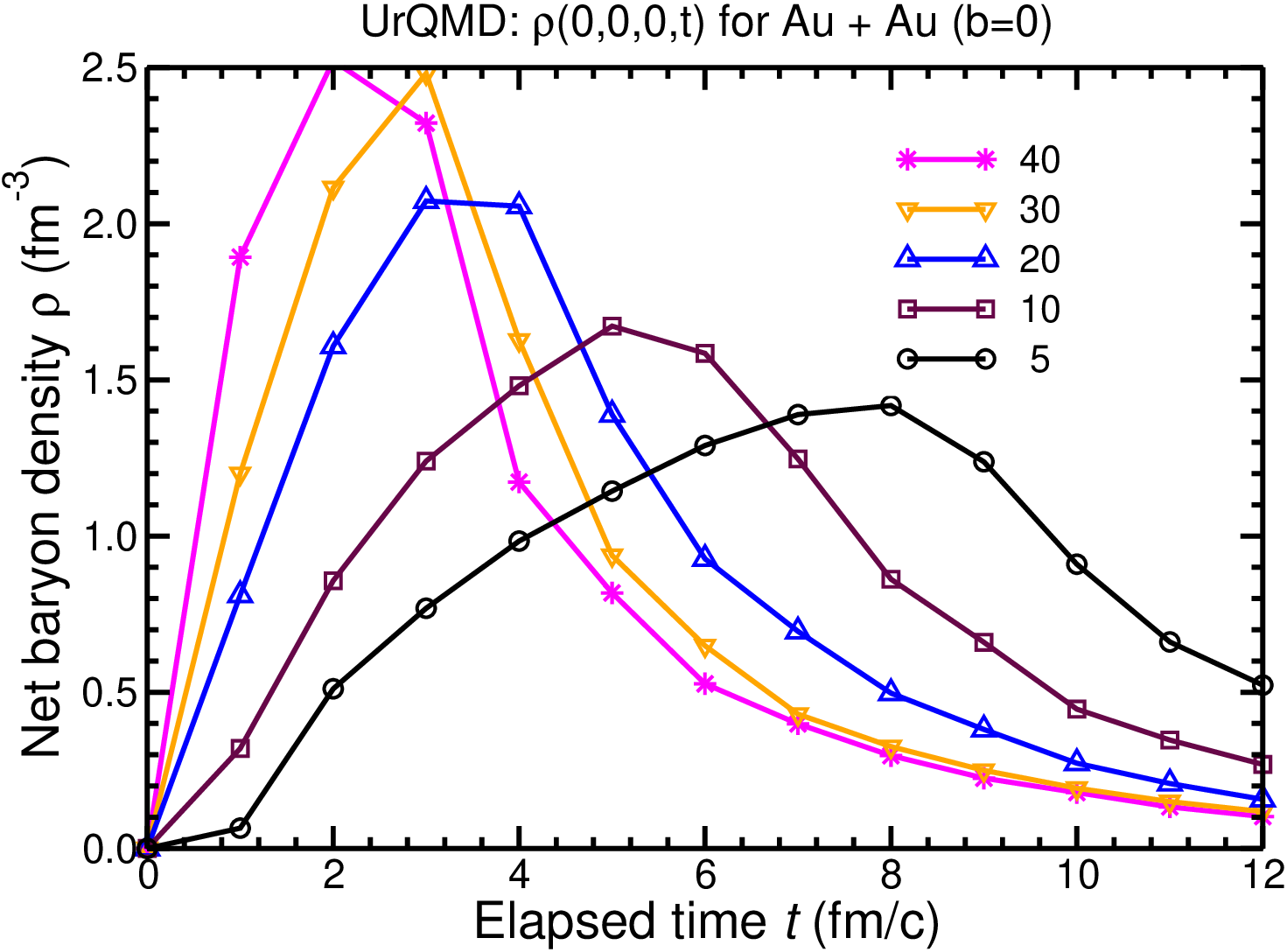}
\includegraphics[angle=0,width=3.1in]{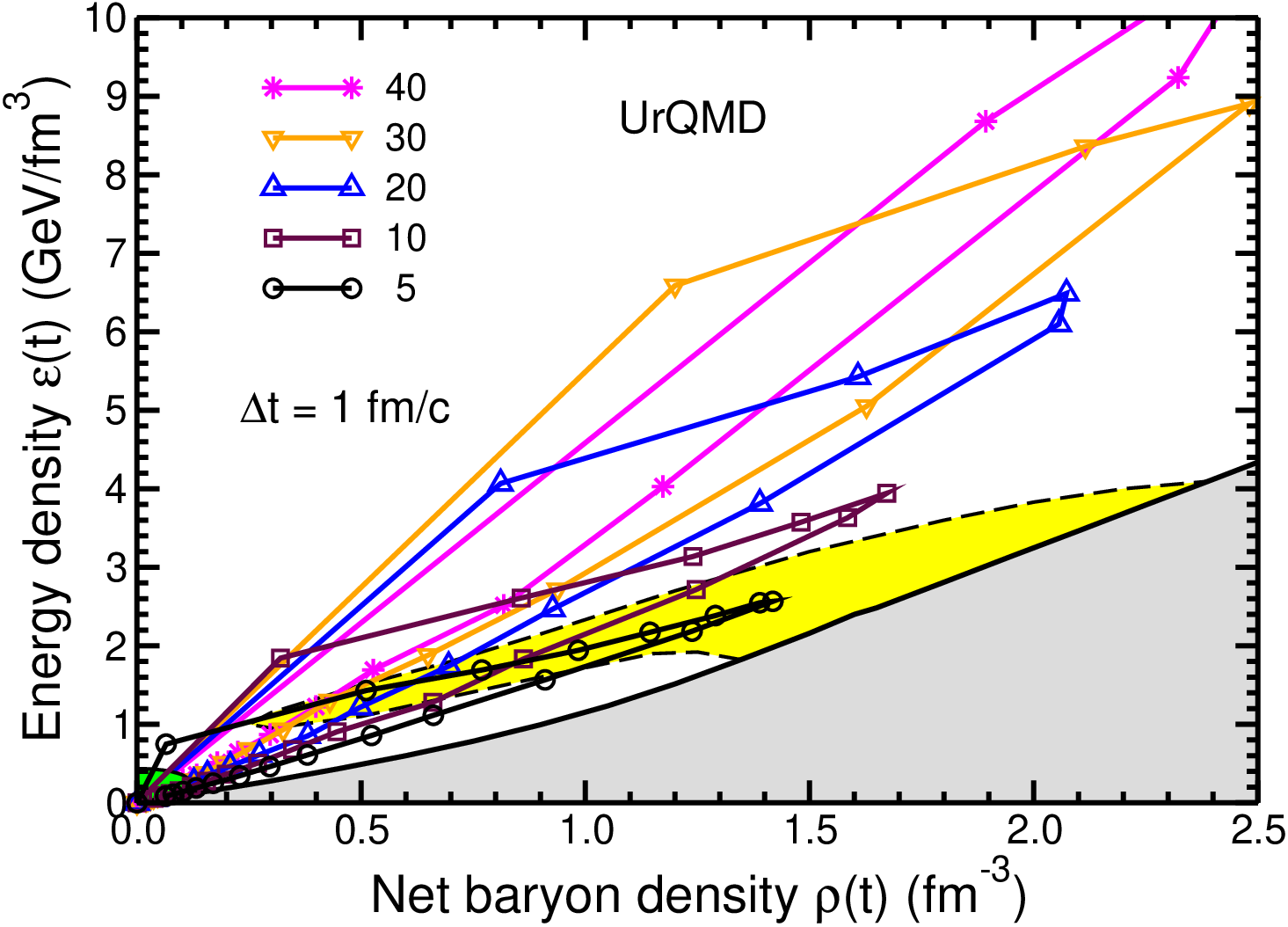}
\caption{[Color online] 
The time evolution of the central net baryon density $\rho(t)$ (top)
and the corresponding phase trajectory $(\rho(t),\eps(t))$ (bottom)
at the center of a head-on Au+Au collision at various bombarding energies
(indicated in GeV/$A$), in the UrQMD model,
together with the schematic reference phase boundary 
depicted in Fig.\ \ref{f:EoS}.
The symbols on the curves are separated by $\Delta t=1~\fm/c$.
}\label{f:UrQMD}
\end{figure}            %       -----------------------------------------

The UrQMD results are shown in Fig.\ \ref{f:UrQMD}.
While the time evolutions of the densities $\rho(t)$ and $\eps(t)$ are 
qualitatively similar to those obtained with the other models discussed,
the UrQMD compressions are somewhat higher than those of PHSD 
and more similar to the QGSM results (see below).
A likely reason for this is that neither UrQMD nor QGSM
has any constraint on the closest approach between two baryons,
whereas both the 3-fluid model and PHSD have some repulsion.

%-----------------------------------------------------------------------
\subsection{Quark-gluon string model}
\label{QGSM}

\begin{figure}[b]          %       -----------------------------------------
\includegraphics[angle=0,width=3.1in]{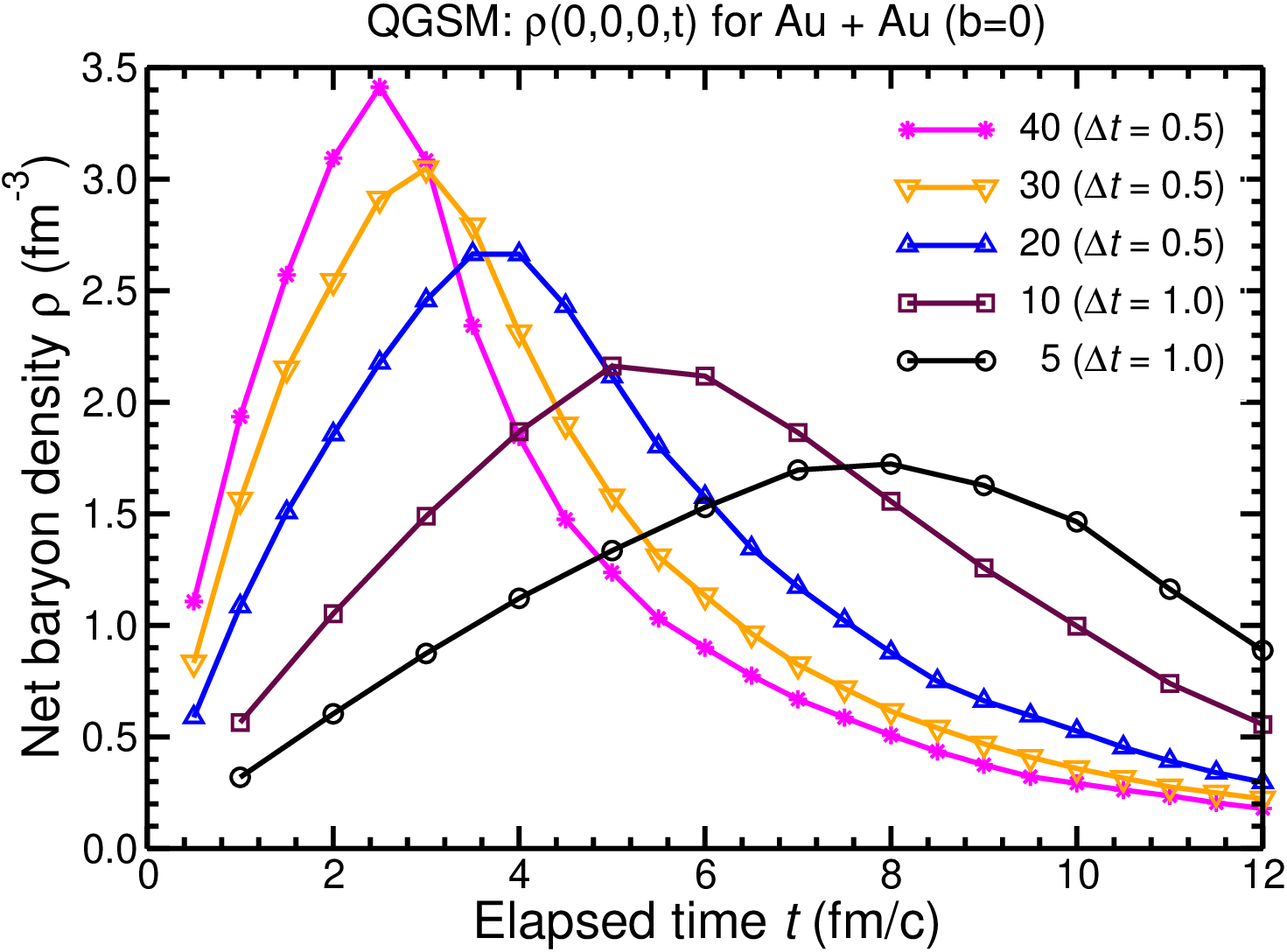}
\includegraphics[angle=0,width=3.1in]{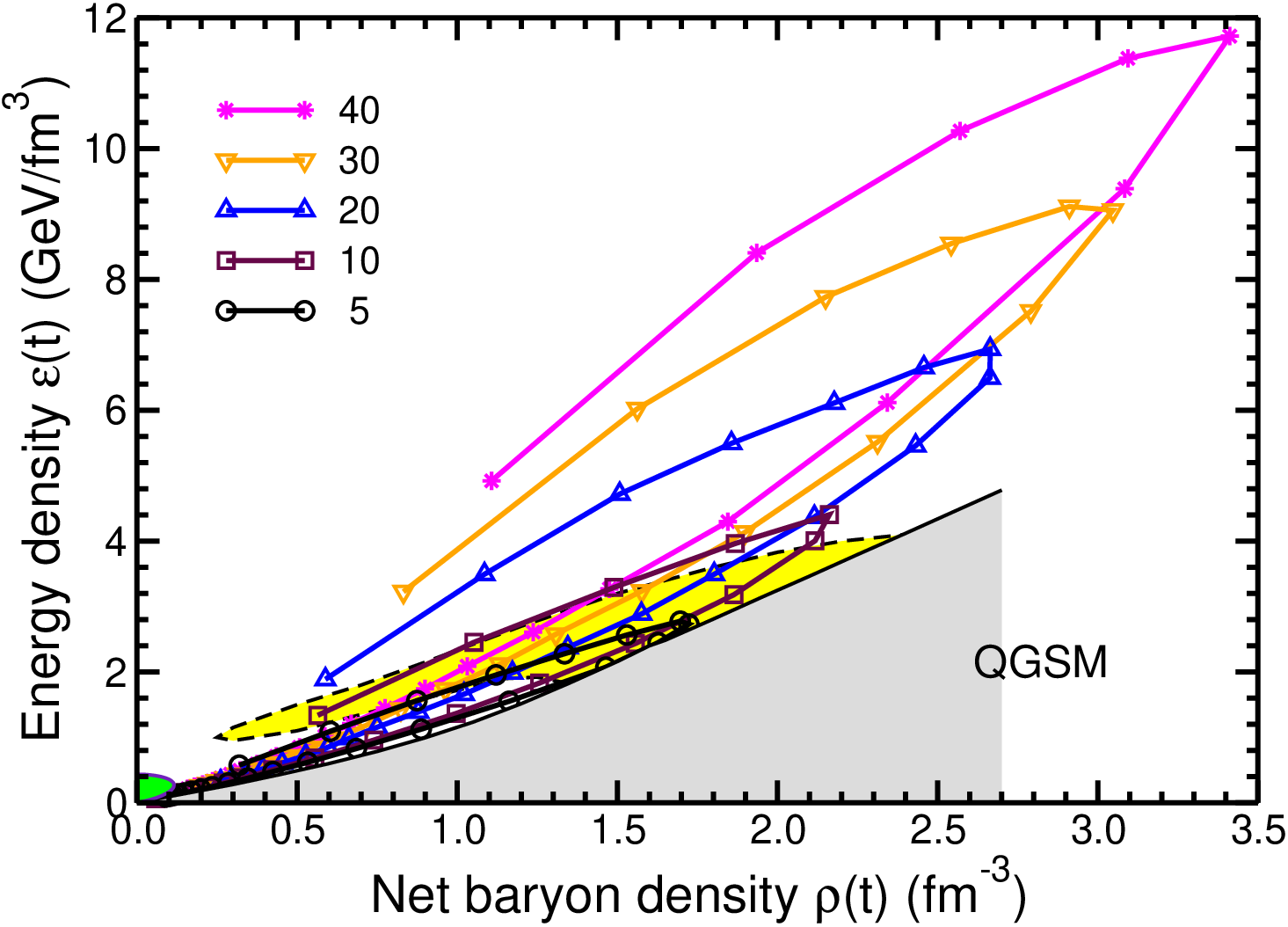}
\caption{[Color online] 
The QGSM evolution of the central net baryon density, $\rho(t)$ (top),
and the corresponding phase trajectory $(\rho(t),\eps(t))$ (bottom)
at the center of a head-on Au+Au collision at various bombarding energies
(indicated in GeV/$A$),
with the time increments $\Delta t$ between the symbols 
indicated in parentheses.
Also shown is the schematic reference phase boundary 
depicted in Fig.\ \ref{f:EoS}.
}\label{f:QGSM}
\end{figure}            %       -----------------------------------------

The Quark-Gluon String Model (QGSM) \cite{Amelin,qgsm2,qgsm3,qgsm4} 
incorporates partonic and hadronic degrees of freedom and is based on 
 Gribov-Regge theory (GRT) \cite{GRT} 
accomplished by a string phenomenology of particle production in inelastic 
hadron-hadron collisions. 
To describe hadron-hadron, hadron-nucleus, and 
nucleus-nucleus collisions, the cascade procedure of multiple secondary 
interactions of hadrons was implemented. 
The QGSM incorporates string fragmentation, formation of resonances, 
and rescattering of hadrons, but simplifies some nuclear effects 
(for example, it neglects the mean fields).

As independent degrees of freedom the QGSM includes 
the octet and decuplet baryons, the octet and nonet vector and pseudoscalar 
mesons, and their antiparticles. 
The initial momenta and positions of nucleons inside 
the nuclei are generated in accordance with the Fermi momentum distribution 
and the Woods-Saxon density distribution, respectively. Pauli blocking of 
occupied final states is taken into account. 

Strings in the QGSM can be produced as a result of 
the colour exchange mechanism or, like in diffractive 
scattering, due to momentum transfer. 
The Pomeron, which is a pole with an intercept $\alpha_P(0) > 1$ in GRT, 
corresponds to the cylinder-type diagrams. 
The inclusive spectra in the 
QGSM have automatically the correct triple-Regge limit for the Feynman 
variable $x \rightarrow 1$, double-Regge limit for $x \rightarrow 0$, and 
satisfy all conservation laws. The particular stages of the collision model, 
namely (i) initialization of interacting projectile and target nuclei, 
(ii) string formation via inelastic nucleon-nucleon (hadron-hadron) 
interaction, (iii) string fragmentation, {\em i.e.}\ hadronization, and 
(iv) hadron-hadron rescattering, are solved basically by Monte Carlo 
simulation techniques.

The results obtained with QGSM are shown in Fig.\ \ref{f:QGSM}.
The time evolution of the densities $\rho(t)$ and $\eps(t)$ are 
qualitatively similar to those obtained with the three models discussed above
and quantitatively closest to UrQMD, 
for the reason explained above (no repulsion).
However, especially at the lower energies,
the QGSM expansion trajectories fall significantly below 
those of the other models.

The model assumptions of UrQMD and QGSM are rather similar
and the absence of hadron repulsion results in higher maximum compressions
than those produced by the other models. 
However, these two models have significantly different expansion dynamics,
probably due to larger number of baryonic resonances included in UrQMD. 
As a result of this, the QGSM dynamics is dominated by the propagation 
of nucleons and pions while UrQMD leads to significant resonance production.
Since the relatively abundant QGSM pions (which are light) 
will propagate at velocities much higher than the UrQMD resonances
(which are heavy), they will leave the interaction region earlier,
thus causing the energy density to decrease more rapidly than in UrQMD.
This mechanism is especially important at the lower energies
where heavy resonances play a significant role
and it gradually subsides as the energy is raised,
consistent with the results.

%-----------------------------------------------------------------------
\subsection{Nuclear Boltzmann equation}

The nuclear Boltzmann equation has proven to be quantitatively useful
for the description of nuclear collisions at lower energies,
up to a few GeV/$A$ and it may therefore be of interest to employ it also here.
For this purpose, we use the Boltzmann-{\"U}hling-Uhlenbeck model
developed by the group in Gie{\ss}en (GiBUU) 
 \cite{EBM99,MEPhD,LM03,WLM05}.

GiBUU explicitly 
propagates 9 $N^*$ and 9 $\Delta$ resonances with masses below 2 GeV
as well as the $S=-1$ hyperons $\Lambda$ and $\Sigma$
and 19 hyperon resonances; the cascades and charmed baryons are included. 
The included mesons are:
$\pi$, $\eta$, $\rho$, $\sigma$, $\omega$, $\eta^\prime$, $\phi$, $\eta_c$, 
$J/\psi$, $K$, $\bar K$, $K^*$, $\bar K^*$. 
The baryon-baryon (meson-baryon)
collisions below $\sqrt{s}=2.6$ (2) GeV are treated within the resonance
scenario, while the string model is applied above.

\begin{figure}          %       -----------------------------------------
\includegraphics[angle=0,width=3.1in]{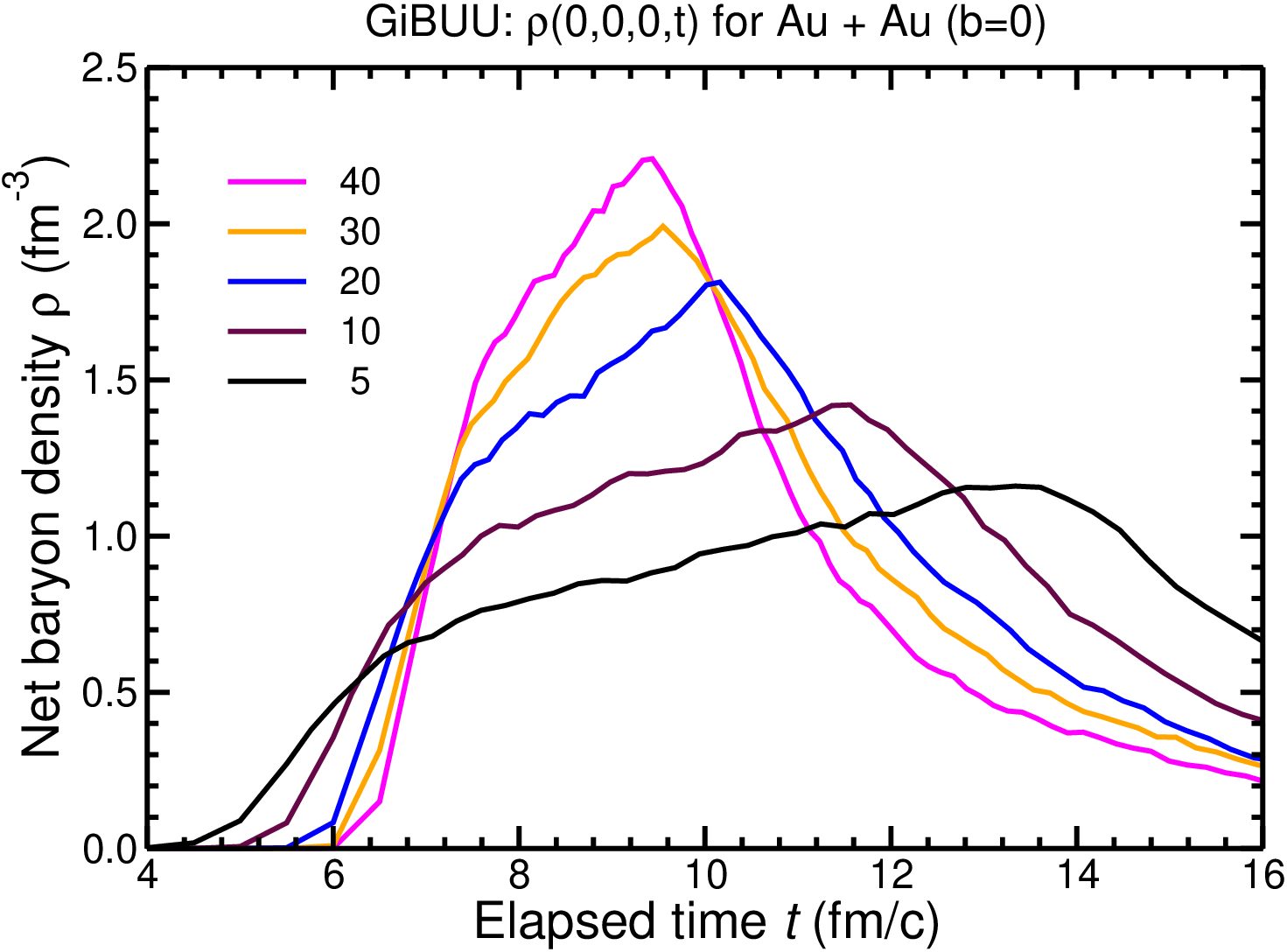}
\includegraphics[angle=0,width=3.1in]{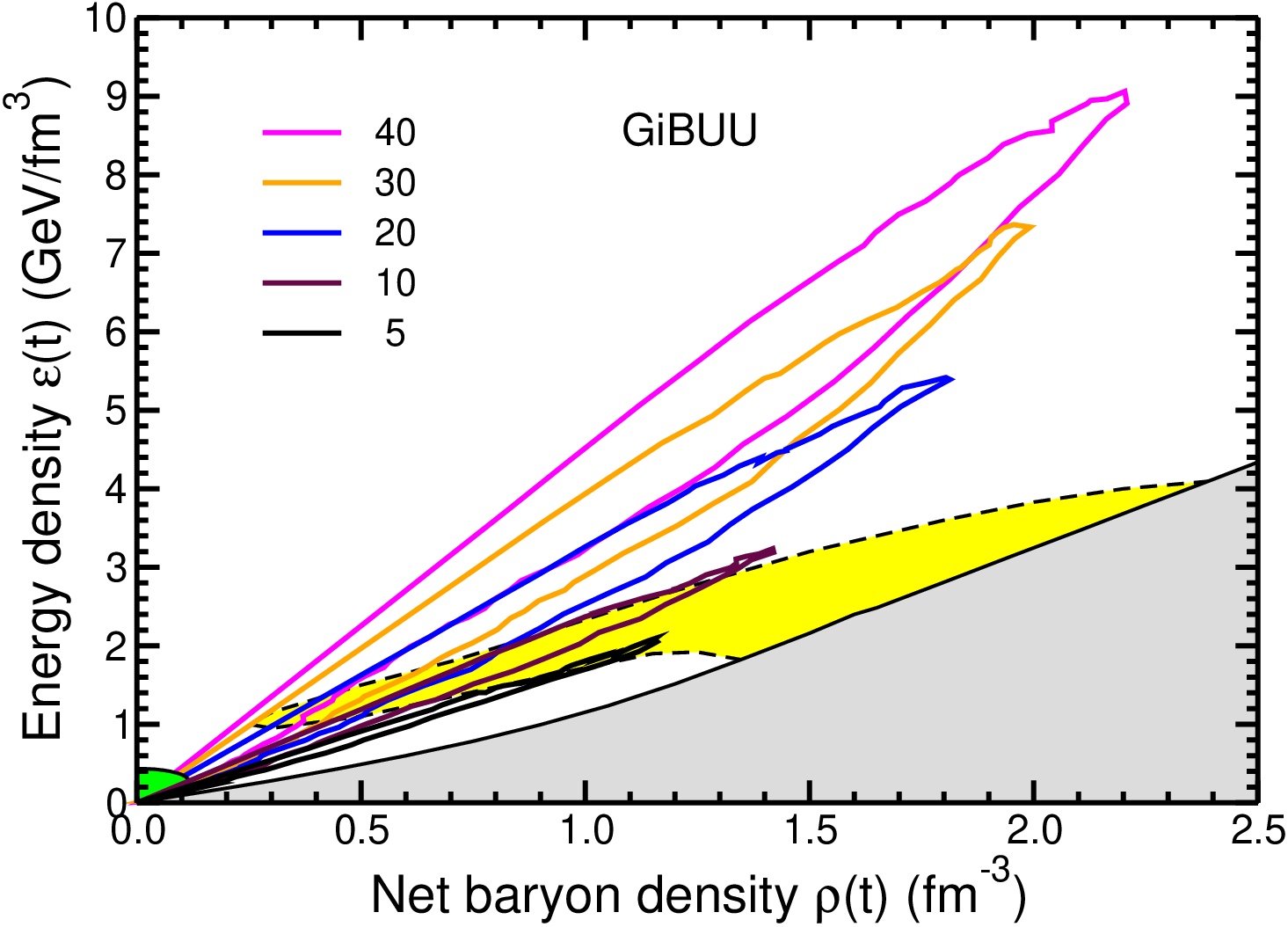}
\caption{[Color online] 
The time evolution of the central net baryon density $\rho(t)$ (top)
and the corresponding phase trajectory $(\rho(t),\eps(t))$ (bottom)
at the center of a head-on Au+Au collision at various bombarding energies
(indicated in GeV/$A$), in the GiBUU model,
together with the schematic phase boundary shown in Fig.\ \ref{f:EoS}.
}\label{f:GiBUU}
\end{figure}            %       -----------------------------------------

Thus GiBUU contains a larger set of the baryonic resonances than most
other transport models (excepting UrQMD and the T\"ubingen QMD model) 
and it consequently  leads to higher pion numbers in vacuum. 
Medium corrections to the cross 
sections $NN \leftrightarrow NR$ and $NN \leftrightarrow NN\pi$ reduce the 
pion number in medium. The in-medium reduced cross sections are implemented 
(optionally) in GiBUU. They are computed with the Dirac masses from the NL2 
model \cite{Lee86}. In particular, the  $NN \leftrightarrow N\Delta$ matrix 
element is given by the one-pion exchange model, as was done in the 
calculations of Dmitriev et al. \cite{DSG86}, but with the vacuum
$\Delta$ and nucleon masses replaced by the Dirac values
which causes a strong in-medium reduction of the cross section \cite{LM03}. 

GiBUU provides a good reproduction of nucleon collective flows \cite{LCGM00} 
as well as pion and kaon multiplicities \cite{LM03,LM05}, at SIS energies.
From AGS to the lower SPS energies, GiBUU overestimates pion multiplicities 
(with vacuum cross sections) but gives a reasonable description of the kaon
multiplicities \cite{WLM05}, as do  HSD and UrQMD.
The direct comparison with the HSD and UrQMD calculations on the pion and kaon
production at 2-40 GeV/$A$ \cite{WLM05} has demonstrated that 
the model yields a somewhat higher $K^+/\pi^+$ ratio 
due to additional meson-meson channels in $K\bar K$ production.

GiBUU is suitable not only for nucleus-nucleus and hadron-nucleus collisions
but also for photon-, electron-, and neutrino-induced reactions. This gives 
the possibility to test the same dynamical part of the model with various 
physical initial conditions. A new numerical realization of the model 
\cite{Buss06} is currently being tested. 

The results presented here are based on the old version 
\cite{EBM99,MEPhD,WLM05}, with the calculations being done in the cascade mode,
{\em i.e.}\ without a mean field and using vacuum cross sections.
These results are shown in Fig.\ \ref{f:GiBUU}.
They are rather similar to those obtained with PHSD,
the main differences being that GiBUU reaches slightly higher densities 
and the difference between the outward and the inward trajectory
grows somewhat faster as the collision energy is raised.

%========================================================================
\section{Comparisons}

\begin{figure}          %       -----------------------------------------
\includegraphics[angle=0,width=3.1in]{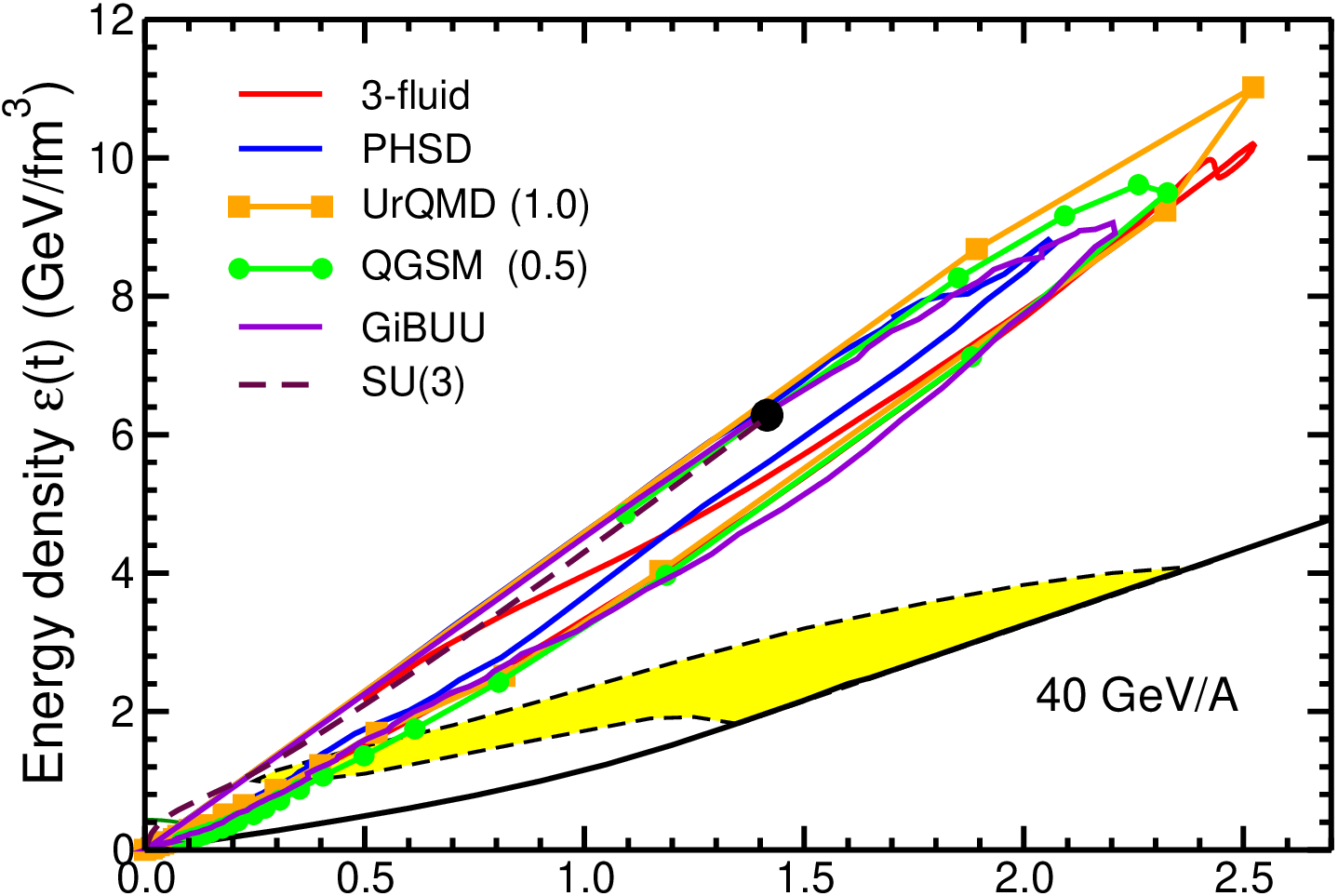}
\includegraphics[angle=0,width=3.1in]{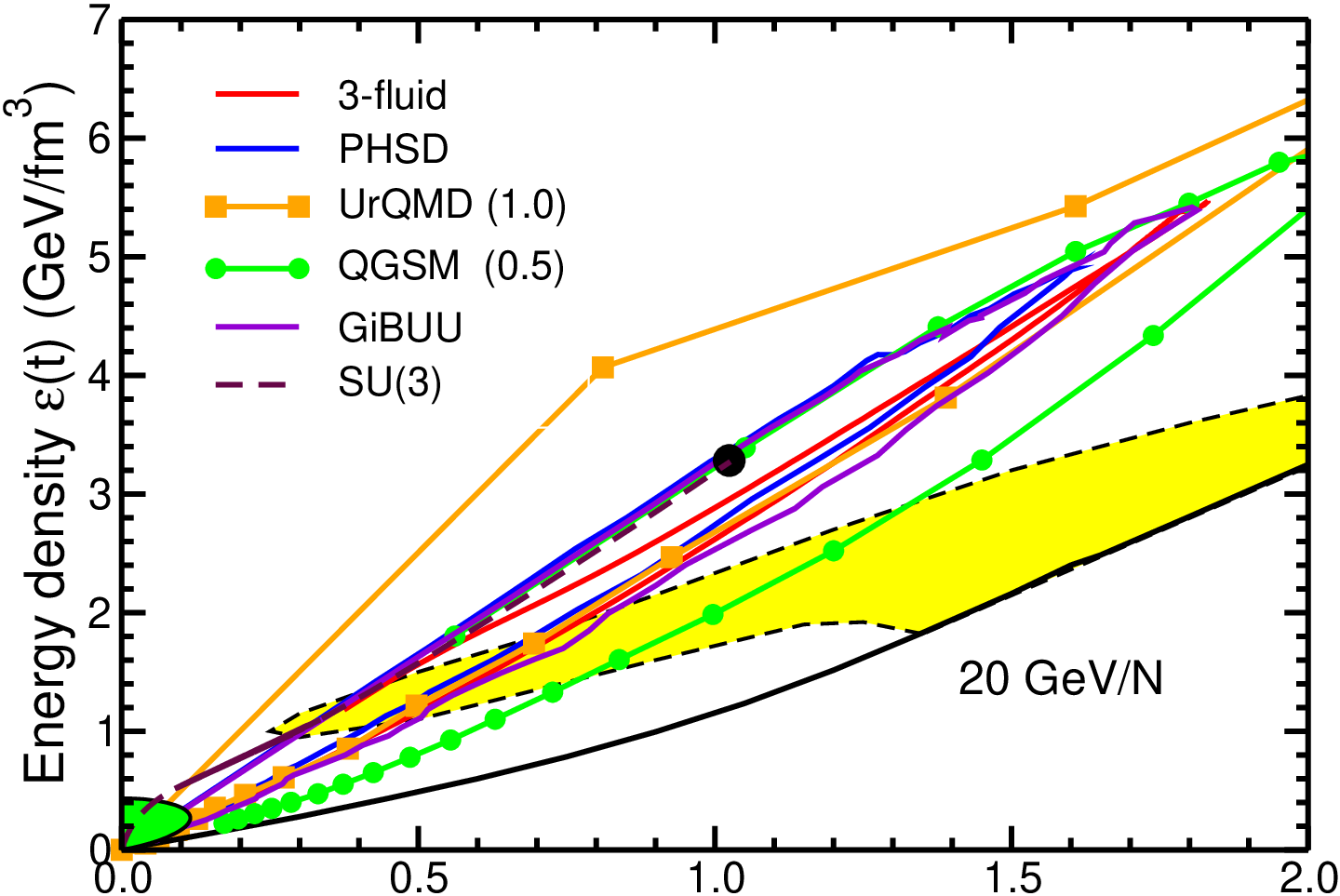}
\includegraphics[angle=0,width=3.1in]{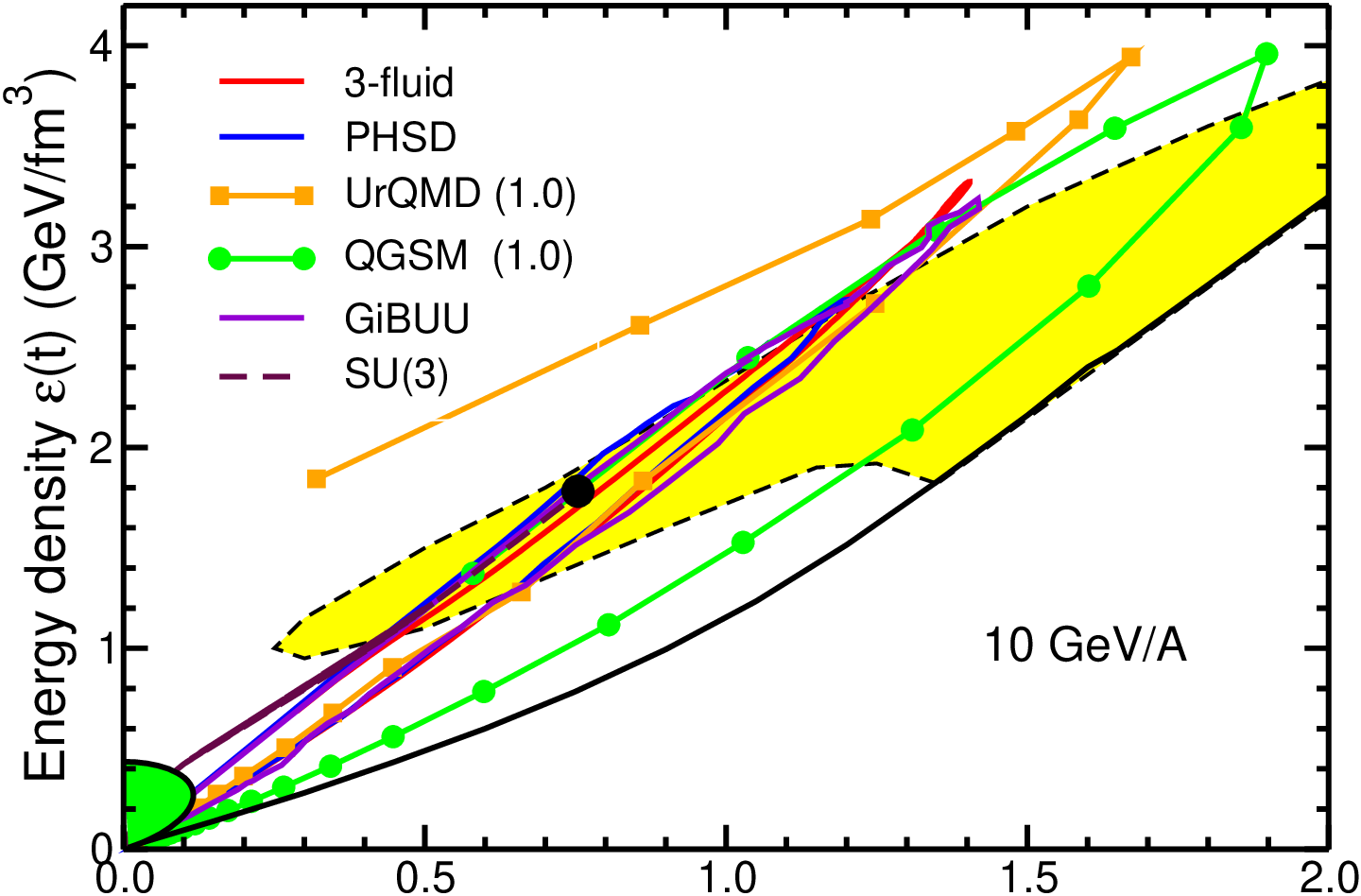}
\includegraphics[angle=0,width=3.1in]{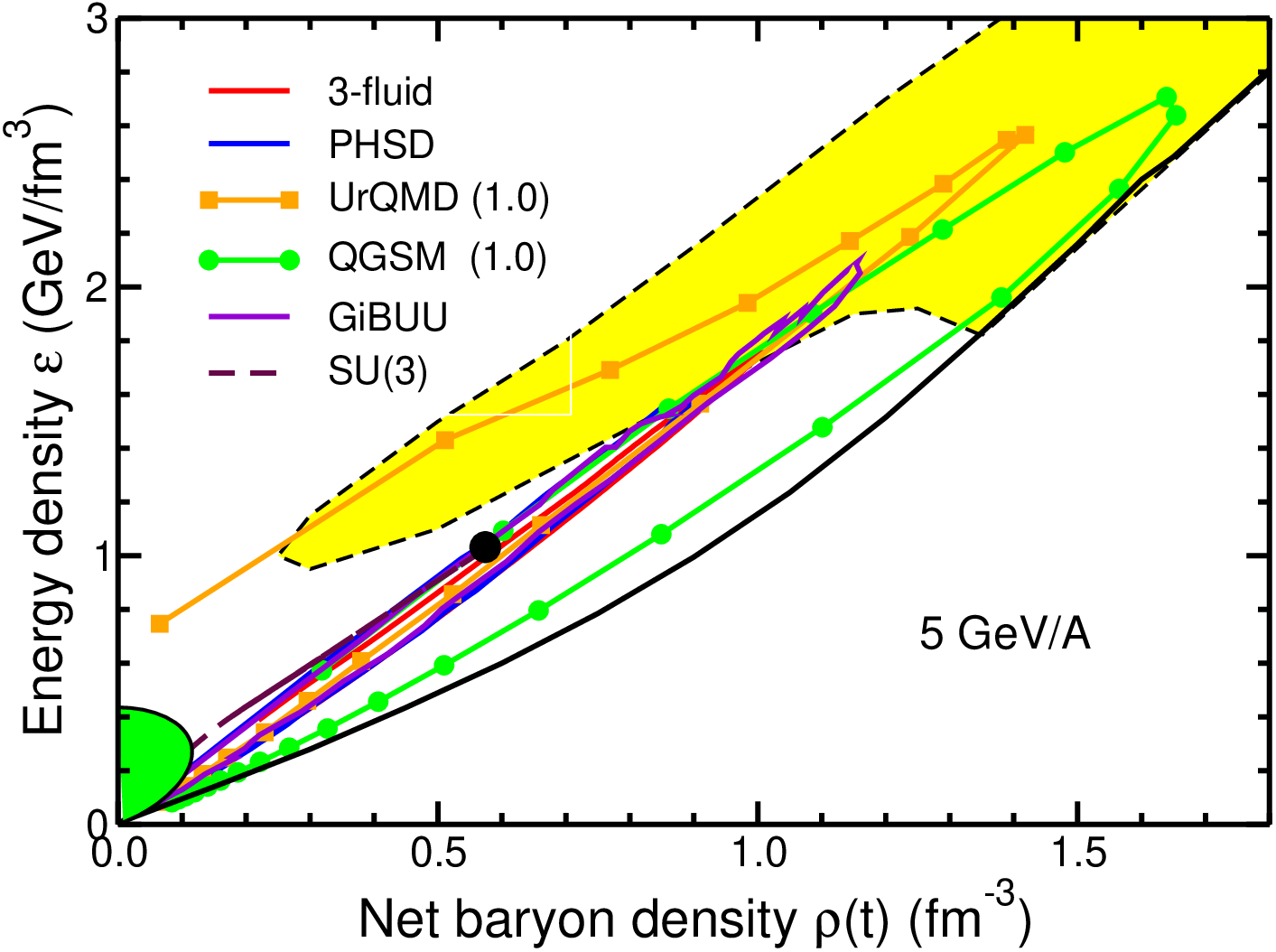}
\caption{[Color online] The phase trajectories $(\rho(t),\eps(t))$
at the center of a head-on Au+Au collision for various bombarding energies
as obtained with the indicated models,
together with the schematic phase boundary shown in Fig.\ \ref{f:EoS}
and the hadronic freeze-out line.
The symbols on the UrQMD and QGSM curves are separated by the indicated time 
intervals.
}\label{f:comp}
\end{figure}            %       -----------------------------------------

We now compare the phase trajectories obtained with the different models
at various collision energies through the anticipated FAIR range.
These comparisons are shown in Fig.\ \ref{f:comp}
for the beam energies $E_0= 5, 10, 20, 40~\GeV/A$.

First of all, we note that by and large there is a remarkable degree of
agreement between the results of the different models.
The most notable exception is the QGSM expansion paths
which come out significantly lower than those of the other models,
as we have already discussed in Sect.\ \ref{QGSM}.

For the subsequent discussion, 
in order to make a concrete analysis possible,
we take the adopted reference phase boundary at face value.
But it is important to keep in mind that this particular boundary,
though not inconsistent with any information we presently have,
is likely to be quantitatively inaccurate.
However, its qualitative form is expected to be correct
and our comments below are therefore expected to be robust,
provided that appropriate adjustments are made 
in the specific energy values mentioned.

At the lowest beam energies 
(under $5~\GeV/A$ or so with the adopted phase boundary),
the degree of compression and agitation attained
does not suffice to bring the central part of the system into the 
phase coexistence region and such collisions are not likely 
to have a bearing on the possible existence of a phase transition.
However, due to their relative slowness,
these collisions achieve of high degree of local equilibrium and
the data obtained in this range may well provide quantitative information 
on the equation of state at the corresponding moderate compressions.

Above those ``subthreshold'' energies follows a range of beam energies
(approximately $5-10~\GeV/A$ for the adopted reference phase boundary),
within which the highest degree of compression occurs 
within the region of phase coexistence.
As the beam energy is increased through this range,
the turning point of the phase trajectory moves across the coexistence region,
starting at the hadronic phase coexistence boundary
and ending at the plasma boundary.
Though somewhat more violent,
these trajectories are generally expected to still attain
a high degree of local equilibrium.
Furthermore, importantly,
they spend the longest period of time within the phase coexistence region.
Therefore, this energy range appears to be especially well suited
for generating signals of the phase transition.

As the collision energy is increased further,
the turning point of the phase trajectory moves further inside
the plasma region and, at the same time, the expansion path steepens.
The time spent crossing the phase coexistence region then decreases,
both in absolute terms and relative to the overall expansion time,
so one would expect any phase-transition signals to gradually subside.

Ultimately, beyond a certain critical collision energy
(for which the expansion phase trajectory 
passes straight through the critical point), 
the phase trajectory never enters the coexistence region
but passes entirely to the left of the critical point.
Though interesting in its own right,
this super-critical region of collision energy would not be expected
to elucidate the character of the deconfinement phase transformation,
{\em i.e.}\ to help determine whether there is in fact a first-order transition
at sufficiently high baryon density.

We also note that the adiabatic expansion results
obtained with the SU(3) model correspond approximately
to the backtracking of the very early (and mostly non-equilibrium) dynamics
when the two Lorentz-contracted nuclear densities are being forced 
to interpenetrate.
This feature mostly reflects the fact that the adopted
initial values were taken to reflect such a scenario.
If suitably modified initial conditions were chosen,
for example obtained from the turning point of a dynamical trajectory,
then the resulting expansion path would exhibit a large degree
of resemblance with the corresponding dynamical trajectory.
Thus one may characterize the actual dynamical expansions
as being approximately adiabatic.

Finally, we wish to emphasize that none of the dynamical models employed
(except possibly PHSD) incorporate a first-order phase transition.
They would therefore not be suitable, in their present form,
for studying actual dynamical consequences of a phase transition.
However, the presence of such a phase transition is not expected to
have an overwhelming effect on the gross dynamics,
primarily due to the predominance of the overall expansion.
[This expectation is supported by comparisons between
HSD (which does not contain a partonic phase) 
and PHSD (which does have a partonic phase) 
in the energy range considered here.]
Therefore, it must also be expected that the effects of a phase transition
would be relatively subtle 
and might best be studied with carefully designed correlation observables.

%========================================================================
\section{Concluding remarks}

The present study has sought to elucidate the bulk conditions
that may be expected to occur in nuclear collisions in the energy range where 
a possible first-order hadronization phase transition would be encountered.
For this we have employed a number of existing dynamical models
to central Au+Au collisions and extracted the time evolution of
net baryon density $\rho$ and the energy density $\eps$ 
at the center of the system 
where these are expected to achieve their largest values.
The different models exhibit a large degree of mutual agreement
on the behavior of the corresponding phase trajectories $(\rho(t),\eps(t))$,
as was summarized in Fig.\ \ref{f:comp},
even if they differ substantially in other regards.

A central issue in the physics of strongly interacting matter
is whether the hadronization phase transformation is of first order
at sufficiently high baryon density.
The calculation of the corresponding critical point,
and the associated phase boundary, poses a significant theoretical challenge 
and the question may ultimately have to be settled by experiment.

The experimental investigation of this question
may employ conceptually different strategies.
One strategy searches for the critical point by means of special signals 
that may occur if the phase trajectory reaches its vicinity.
However, the location of the critical point remains theoretically
poorly understood and, as is well illustrated by Fig.\ \ref{f:comp},
a small shift in its position
would require a relatively large change in the collision energy
of the critical phase trajectory that passes through it.
Consequently, it is hard to predict what energy range 
would be most suitable for this approach.
Indeed, our present studies cannot rule out 
that the critical collision energy lies somewhere in the SPS range
above the energies reachable by the planned FAIR.

A different experimental strategy seeks direct evidence 
of the first-order transition by concentrating on signals that might appear 
as a result of the phase trajectory encountering the phase-transition line.
One would expect that such signals would best be generated 
if the bulk of the system were brought well inside 
the phase coexistence region, where a phase decomposition is favored,
and kept there for a time sufficient to allow the development of
the macroscopic non-uniformities associated with the phase decomposition.
Though also associated with significant uncertainties,
this issue can probably be assessed with somewhat larger confidence.
Thus, considering the large degree of mutual agreement between the
different dynamical results and
taking first the adopted schematic phase boundary at face value,
the present study suggest that the optimal beam energy is around $10~\GeV/A$,
corresponding to $\sqrt{s_{NN}} \approx 2.36~\GeV + 2.36~\GeV$ for a collider.
We must, however, make allowance for the fact that the adopted schematic
phase boundary is not expected to be quantitatively accurate.
Furthermore, the dynamical models, though yielding fairly similar results,
may all possess common inaccuracies.
With a factor of two admitted to account for such uncertainties,
the present study would then suggest that the optimal conditions for exploring 
the hadronization phase transition are likely (though not certain)
to occur for beam kinetic energies of $5-20~\GeV/A$,
corresponding to $\sqrt{s_{NN}} \approx 3.6-6.4~\GeV$ for a collider.

These numbers suggest that the planned FAIR at GSI 
is well matched for such studies.  Furthermore,
this region may also be accessible at the low-energy end of RHIC at BNL,
as well as at a possible upgraded Nuclotron at JINR in Dubna.
Of course, further dynamical studies are required before it is possible
to identify the specific candidate signals and to assess whether they
can indeed be expected to develop sufficiently even at the optimal
collision energy.  
We hope that this study will provide stimulation in this regard.
\\~\\

%========================================================================
\section*{Acknowledgments}

This work was supported by the Office of Energy Research
%Office of High Energy and Nuclear Physics, Division of Nuclear Physics,
%the Office of Basic Energy Science, Division of Nuclear Science, 
at the U.S.\ Department of Energy (DOE Contract DE-AC03-76SF00098),
the Deutsche Forschungsgemainschaft (DFG project 436 RUS 113/558/0-3), 
and the Russian Foundation for Basic Research (RFBR grant 06-02-04001 NNIO\_a),
and Russian Federal Agency for Science and Innovations (grant NSh-8756.2006.2).
The authors appreciate the hospitality of the 
Gesellschaft of Schwerionenforschung (GSI) in Darmstadt, where part of this
work was carried out,
and also acknowledge helpful discussions with K.\ Gudima.

%========================================================================
{}
%========================================================================

			\end{document}